\documentclass{article}
\usepackage[hyphens]{url}
\usepackage[utf8]{inputenc}
\usepackage{graphicx}
\usepackage{authblk}
\usepackage[english]{babel}
\usepackage{setspace}
\usepackage{fancyhdr}
\usepackage{array}
\usepackage[margin=1in]{geometry}
\usepackage{float}
\usepackage{amsmath}
\usepackage{amssymb}
\usepackage{bbm}
\usepackage{bm}
\usepackage{scalerel,stackengine}
\usepackage[final]{pdfpages}
\usepackage{longtable}
\usepackage{chngcntr}
\usepackage{placeins}
\usepackage{microtype}
\usepackage{tikz}
\usepackage{makecell}
\usepackage{hyperref}
\usepackage{subcaption}
\usepackage{rotating}
\usepackage{soul}
\usepackage{tabularx} 
\usepackage{pdfpages}
\usepackage[square,numbers]{natbib}
\usepackage{multirow}

\hypersetup{
    colorlinks = true,
    citecolor = {blue},
    urlcolor = {blue},
    menucolor = {blue},
    linkcolor = {blue}
}

\makeatletter
\patchcmd{\@maketitle}{\LARGE \@title}{\fontsize{18}{20}\selectfont\textbf{\@title}}{}{}
\makeatother

\title{Adaptive clinical trial designs with blinded selection of binary composite endpoints and sample size reassessment}  

\author[1]{Marta Bofill Roig\thanks{marta.bofillroig@meduniwien.ac.at}} 
\author[2]{Guadalupe G\'{o}mez Melis}  
\author[1]{Martin Posch}
\author[1]{Franz Koenig}

\affil[1]{Center for Medical Statistics, Informatics and Intelligent Systems, Medical University of Vienna, Vienna, Austria}
\affil[2]{Departament d'Estad\'{i}stica i Investigaci\'{o} Operativa, Universitat Polit\`{e}cnica de Catalunya-BarcelonaTECH, Barcelona, Spain}

\date{}         
\begin{document}

\maketitle

\begin{abstract}
	{For randomized clinical trials where a single, primary, binary endpoint would require unfeasibly large sample sizes, composite endpoints are widely chosen as the primary endpoint. Despite being commonly used, composite endpoints entail challenges in designing and interpreting results. Given that the components may be of different relevance and have different effect sizes, the choice of components must be made carefully. Especially, sample size calculations for composite binary endpoints depend not only on the anticipated effect sizes and event probabilities of the composite components, but also on the correlation between them. However, information on the correlation between endpoints is usually not reported in the literature which can be an obstacle for planning of future sound trial design. 
		We consider two-arm randomized controlled trials with a primary composite binary endpoint and an endpoint that consists only of the clinically more important component of the composite endpoint. We propose a trial design that allows an adaptive modification of the primary endpoint based on blinded information obtained at an interim analysis. Especially, we consider a decision rule to select between a composite endpoint and its most relevant component as primary endpoint. The decision rule chooses the endpoint with the lower estimated required sample size. Additionally, the sample size is reassessed using the estimated event probabilities and correlation, and the expected effect sizes of the composite components. We investigate the statistical power and significance level under the proposed design through simulations. We show that the adaptive design is equally or more powerful than designs without adaptive modification on the primary endpoint. Besides, the targeted power is achieved even if the correlation is misspecified at the planning stage while maintaining the type 1 error. All the computations are implemented in R and illustrated by means of a peritoneal dialysis trial.}
	{Clinical Trial; Composite Endpoint; Interim Analysis; Adaptive design; Sample Size.}
\end{abstract}

\section{Introduction} 
\label{sect_intro}

Composite endpoints are frequently used in randomized controlled trials (RCTs) to provide a more comprehensive characterization of patients’ response than when using a single endpoint. For example, 
Major Adverse Cardiovascular Events 
in cardiovascular disease, where the composite endpoint includes death, stroke, myocardial infarction, or revascularization is commonly used for time-to-event endpoints \citep{gerstein2021cardiovascular} and binary endpoints \citep{Cordoba2010}. 
The use of composite endpoints can also improve the power in situations where the incidence rates of the individual components are too low to achieve adequate power with feasible sample sizes and trial durations. The combination of several components into a composite endpoint provides then a solution by increasing the incidence rate of the primary endpoint.  
However, using composite endpoints comes with a cost. The interpretation becomes more complex, especially when components have different effect sizes and different event probabilities. Moreover, if the treatment has only an effect in some components, the effect size of the composite will be diluted. 
When a composite endpoint is used as primary endpoint, regulatory agencies require to analyse in addition all components separately as secondary endpoints \citep{FDA_2017, EMA_2017, Mao2021}. In particular, it is necessary to assess the effects of the most relevant component under study.
When designing a trial with a composite endpoint, sample size calculation is especially challenging since it requires the anticipation of event probabilities and effect sizes of the components of the composite endpoint as well as the correlation between them. While the marginal effect size of each component are usually known, the correlation is often not reported. 

In the context of peritoneal dialysis, the binary composite endpoint major adverse peritoneal events (MAPE) has been recently proposed \citep{Boehm2019}. 
This endpoint combines three individual components: 
(1) peritonitis, (2) peritoneal membrane deterioration, and (3) technical failure;  where peritonitis and peritonitis membrane deterioration endpoints are considered clinically more relevant. Given that this composite endpoint is relatively new, there is only limited data as basis for sample size calculations available. 
So under which circumstances is it best to consider the composite endpoint MAPE in terms of power of the trial? Or how could we design the trial robustly to possible deviations from the anticipated correlation? In this work, we aim at addressing both questions. We propose a design in which the decision of  whether it is better to consider the composite endpoint or its most relevant component as the primary endpoint is reevaluated by choosing the endpoint with the smaller required sample size. Based on this choice,  the sample size is recalculated, incorporating correlation information estimated at an interim analysis if necessary. Adaptations to endpoint selection and, in particular, designs that allow adaptive modification of the primary endpoint based on interim results are discussed in the Food and Drug Administration guidance on adaptive designs \citep{FDA_2017,FDA_2019}. Regulatory agencies require the adaptation rule to be planned before the data become available and the use of appropriate statistical methods to ensure that the type 1 error is controlled.

In trials with multiple endpoints of interest, the testing strategy can either be based on a single endpoint (and thus consider the rest as secondary endpoints), combining all the endpoints in a composite endpoint, or considering a multiple test using all the endpoints. 
The choice of the primary composite endpoint based on the trial's efficiency has been addressed by several authors. 
\citet{Lefkopoulou1993} compared the use of multiple primary endpoints to a composite endpoint by means of the Asymptotic Relative Efficiency (ARE) between the corresponding hypothesis tests. 
\citet{Gomez2013a} 
and  \citet{BofillRoig2018} 
proposed the ARE as a method to choose between a composite endpoint or one of its components as primary endpoint for comparing the efficacy of a treatment against a control in trials with survival data and binary data, respectively. 
\citet{Sozu2016} evaluated the efficiency of the trial depending on the number of endpoints considered.

Several authors have proposed different approaches to size trials with several endpoints as primary. \citet{Sozu2010} 
discussed sample size formulae for multiple binary endpoints. As it is known,  a major difficulty in the sample size calculation is that sometimes the required information depends on nuisance parameters or  highly variable parameters.  
In trials with multiple endpoints, the required sample size depends on the correlation among the considered endpoints and needs to be taken into account in sample size calculations \citep{FDA_2017,EMA_2017}. 
However, the correlation between endpoints is usually unknown and often not reported in the literature which can be an obstacle for sound trial design. 
Several authors showed, that the correlation has a large impact on the required sample size when using multiple co-primary and composite binary endpoints \citep{Sozu2010, Bofill2019}.
One way to address this problem may be to consider an interim analysis to estimate unknown parameters, in particular, the correlation. Existing work in this context has mainly focused on trials with multiple endpoints. 
\cite{Kunz2017} approached the sample size calculation of trials with multiple, correlated endpoints. They proposed  estimators for the covariance and the correlation based on blinded data obtained at an interim analysis. 
\cite{Sander2017} considered trials in which the composite endpoint and its most relevant component are two primary endpoints. They proposed an internal pilot study design where the correlation between the statistics for the composite endpoint and the most relevant component is estimated in a blinded way at an interim stage and where the sample size is then revised accordingly. 
Surprisingly, less attention has been given to the estimation of the correlation between the components of composite endpoints per se and sample size reassessment in trials with primary  composite endpoints. 

In this paper, we propose a trial design that allows an adaptive modification of the primary endpoint based on blinded information obtained at an interim analysis and recalculates the sample size accordingly. If the primary endpoint is decided to be the composite endpoint, then the sample size reassessment incorporates the information of the estimated correlation. 
We focus on a two-arm RCT with a primary composite binary endpoint defined by two components, of which one is considered clinically more relevant. 
In Section \ref{sect_notation}, we present the problem setting and our main objectives. In Section \ref{sect_adapdesign}, we propose the adaptive design with endpoint modification. We first introduce the decision rule used to adaptively select the primary endpoint. Then we discuss how this decision rule is computed based on blinded data and the subsequent sample size recalculation. In Section \ref{sect_extension}, we extend the proposed design for trials with composite endpoints of more than two components and more than two arms. 
In Section \ref{sect_example} we apply our methods to Peritoneal Dialysis trials. Furthermore, in the supplementary material, we present an R package in which the methodology has been implemented and include an additional example in the context of cardiology trials in which the R code is provided as a tutorial. We performed a blinded selection of the primary endpoint using the observed data from a conducted trial. In Section \ref{sect_sim}, we evaluate the operating characteristics of the adaptive design. We finish with a short discussion. 

The R code to implement the proposed methods and reproduce the results of this article is available at \url{https://github.com/MartaBofillRoig/eselect}.

\section{Notation, hypotheses and trial designs}
\label{sect_notation}

Consider an RCT designed to compare two treatment groups, a control group ($i=0$) and an intervention group ($i=1$), each composed of $n^{(i)}$ individuals, and denoting by $n=n^{(0)}+n^{(1)}$ the total sample size and by $\pi=n^{(0)}/n$ the allocation proportion to the control group. 
Assume two events of interest, say $\varepsilon_1$ and $\varepsilon_2$, and assume that there is one event (say $\varepsilon_1$) which is more relevant for the scientific question than the other.  
Let $X_{ijk}$ denote the response of the $k$-th binary endpoint for the $j$-th patient in the $i$-th group ($i=0,1$, $j=1,...,n^{(i)}$, $k=1,2$).
The response $X_{ijk}$ is $1$ if the event $\varepsilon_k$ has occurred during the follow-up and $0$ otherwise. 
Let $p_k^{(i)}$ represent the probability that $\varepsilon_k$ occurs for a patient belonging to the $i$-th group. 
Let 
$\textrm{OR}_k = \frac{p_k^{(1)}/q_k^{(1)}}{p_k^{(0)}/q_k^{(0)}}$ denote the odds ratio for the $k$-th endpoint, where $q_k^{(i)}=1-p_k^{(i)}$ ($i=0,1,k=1,2$).

Define the composite binary endpoint as the event that occurs whenever one of the endpoints $\varepsilon_1$ and $\varepsilon_2$ is observed, that is, $\varepsilon_* = \varepsilon_1 \cup \varepsilon_2$. Denote by $X_{ij*}$ the composite response defined as:
\begin{eqnarray*} 
	X_{ij*} &=&
	\begin{cases}
		1, \text{ if } X_{ij1} + X_{ij2} \geq 1 \\
		0, \text{ otherwise}.
	\end{cases}
\end{eqnarray*}
Let $p^{(i)}_{*}$ be the event probability of the composite endpoint, $p^{(i)}_{*}=\mathrm{P}(X_{ij*} =1)$, and $OR_*$ be the odds ratio for the composite endpoint $\varepsilon_*$. 
We denote by $\hat{p}_k^{(i)}$ the estimated probability of response for the $k$-th binary endpoint in group $i$, that is, 
$\hat{p}_k^{(i)} = \frac{1}{n^{(i)}} \sum_{j=1}^{n^{(i)}} X_{ijk} = 1 - 	\hat{q}_k^{(i)}$.

\subsection{Trial design using the composite endpoint}\label{design_CE}

Assume that initially the trial is planned with the composite endpoint $\varepsilon_* = \varepsilon_1 \cup \varepsilon_2$ as the primary endpoint. The hypothesis to be tested is  the null hypothesis of no treatment difference in the composite endpoint $H_*: OR_*=1$ against the alternative hypothesis of a risk reduction in the treatment group, $K_*: OR_*<1$. 
We test $H_*$ using the test statistic $T_{*,n}$, given by: 
\begin{eqnarray} \label{test:ch} 
	T_{*,n} &=& \frac{ \log(\widehat{OR}_*) }{
		\sqrt{ \frac{1}{n^{(0)}\hat{p}_*^{(0)} \hat{q}_*^{(0)} } + \frac{1}{n^{(1)}\hat{p}_*^{(1)} \hat{q}_*^{(1)}} }} 
\end{eqnarray} 
This statistic is asymptotically $N(0,1)$ under $H_*$ and we reject the null hypothesis if $T_{*,n}<z_\alpha$, where $z_x$ denotes the quantile of the standard normal distribution \citep{chow2017sample}. 
Then the sample size needed to achieve a power of $1-\beta$ given a significance level $\alpha$ is
\begin{eqnarray} \label{ss_CBE}
	N_*(p_*^{(0)},OR_*) &=& \left( \frac{z_\alpha+z_\beta}{\log(OR_*)} \right)^2 \left( \frac{1}{p_*^{(0)} (1-p_*^{(0)})} + \frac{1}{\left(\frac{1-\pi}{\pi} \right) p_*^{(1)} (1-p_*^{(1)})} \right).
\end{eqnarray}

Thus, to size a trial with  a composite endpoint as primary endpoint, we need to specify the probability of an event in the composite endpoint in the control group and the odds ratio. 
If information on the parameters of the joint distribution of the components is available, 
the distribution of the composite endpoint can be derived (\cite{Bofill2019}). 
Specifically, the event probability of the composite endpoint in the $i$-th group, $p_*^{(i)}$, is determined by the probabilities of the components, $p_1^{(i)}$ and $p_2^{(i)}$, and Pearson's correlation coefficient between the components, $\rho$, as follows:
\begin{eqnarray} \label{prob_CBE} 
	p^{(i)}_{*} &=& 1- q_1^{(i)} q_2^{(i)} - \rho \sqrt{p_1^{(i)} p_2^{(i)} q_1^{(i)} q_2^{(i)}} 
\end{eqnarray}
The odds ratio for the composite endpoint, $OR_*=\mathrm{OR}_*(p_1^{(0)},p_2^{(0)},OR_1,OR_2,\rho)$, can be expressed as function of the odds ratios $OR_1, OR_2$, the event probabilities in the control group, $p_1^{(0)}, p_2^{(0)}$, and the correlation $\rho$ (see the  supplementary material).  
Note, however, that in both cases, to compute  $p^{(i)}_{*}$ (in \eqref{prob_CBE}) and $\mathrm{OR}_*(p_1^{(0)},p_2^{(0)},OR_1,OR_2,\rho)$, we make the underlying assumption that the correlation between the components is the same in the treatment and control groups. Although we focus on the correlation in this work, other association measures can be used instead. In the supplementary material, we present different association measures, such as the relative overlap and conditional probability, and establish the relationship between them and the correlation so that one can move from one to the other depending on what is easier to anticipate. Further details regarding the assumption of equal correlations across arms can be found in the supplementary material.

As a consequence, the required sample size $N_*(p_*^{(0)},OR_*)$ can be computed based on $p_*^{(0)}$, given in \eqref{prob_CBE}, and $\mathrm{OR}_*$, given in equation (1) in the supplementary material. With a slight abuse of notation, we refer to the sample size computed by means of the components' parameters as $N_*(p_1^{(0)},p_2^{(0)},OR_1,OR_2,\rho)$.

\subsection{Trial design using the most relevant endpoint only}\label{design_RE}
The null and alternative hypotheses related to the most relevant of the components, $\varepsilon_1$, are $H_{1}: OR_1 = 1$ and $K_{1}: OR_1<1$. 
Similar to the composite design, let $T_{1,n}$ be the statistic to test $H_1$, defined by 
\begin{eqnarray} \label{test:eh} 
	T_{1,n} &=& \frac{ \log(\widehat{OR}_1) }{
		\sqrt{ \frac{1}{n^{(0)}\hat{p}_1^{(0)} \hat{q}_1^{(0)} } + \frac{1}{n^{(1)}\hat{p}_1^{(1)} \hat{q}_1^{(1)}} }}.
\end{eqnarray} 
As above, $T_{1,n}$ is asymptotically $N(0,1)$ under $H_1$, and the null hypothesis $H_{1}$ is rejected  if $T_{1,n}<z_\alpha$.  
The sample size $N_1(p_1^{(0)}, OR_1)$ required to achieve a power of $1-\beta$ at a one-sided significance level of $\alpha$ is given by \eqref{ss_CBE} replacing $p_1^{(0)}$ and $OR_1$ by $p_*^{(0)}$ and $OR_*$, respectively. 


\section{Adaptive design with endpoint modification}
\label{sect_adapdesign}

\subsection{Decision rule based on the ratio of sample sizes} 

We propose a trial design that allows modifying adaptively the primary endpoint based on blinded information obtained at an interim analysis or at the end of the trial. 
The decision rule to select the endpoint to be used as the primary endpoint chooses the endpoint with the lower estimated required sample size. Let $d(\cdot)$ denote the ratio of the required sample size for each of the designs, given by
\begin{eqnarray} \label{decision}
	d(p_1^{(0)},p_2^{(0)},OR_1,OR_2,\rho) &=& \frac{N_1(p_1^{(0)},OR_1)}{N_*(p_1^{(0)},p_2^{(0)},OR_1,OR_2,\rho)} 
\end{eqnarray}
where $N_1(\cdot)$ and $N_*(\cdot)$ are the sample sizes for the relevant and composite endpoints introduced in Subsections \ref{design_CE} and \ref{design_RE}, respectively. Note that this ratio depends also on  $\alpha$ and $\beta$.
Now, the decision rule to select the primary endpoint is as follows: If $d(\cdot)<1$,  use the most relevant endpoint as the primary endpoint; if $d(\cdot)\geq 1$ the composite endpoint is chosen. 

\subsection{Estimation of the sample size ratio based on blinded data}\label{DecisionRule}

In order to estimate the sample size ratio of the designs with the most relevant and the composite endpoint, we use the blinded data obtained either at the interim analysis or the end of the trial. Specifically, we derive estimates of the event probabilities of the components in the control group and their correlation. Besides the blinded (interim) data, the estimates are based on the a priori assumptions on the effect sizes.

Suppose that the blinded analysis, using the pooled sample, is based on a sample of size $\tilde{n}$, where $\tilde{n}$ could be the total sample size initially planned ($\tilde{n}=n$) or a proportion of it used at an interim stage ($\tilde{n}=\omega \cdot n$, with $0<\omega<1$). Also, suppose that the proportion of patients assigned to the control group based on this sample is the same as the one expected at the end of the trial, that is, $\pi=n^{(0)}/n=\tilde{n}^{(0)}/\tilde{n}$, where $\tilde{n}^{(0)}$ is the sample size in the control group in the blinded data. 
Based on the observed responses in the pooled sample, we estimate the probabilities $p_1$, $p_2$, and $p_*$, where $p_k=\pi p_k^{(0)} + (1-\pi)p_k^{(1)}$ for $k=1,2,*$ and $\pi=n^{(0)}/n$.
Assuming that the expected effects for the components ($OR_1$ and $OR_2$) have been pre-specified in advance, we obtain estimates of the probabilities of each composite component under the control group $p_1^{(0)},p_2^{(0)}$ and subsequently the estimates of the probabilities under the treatment group $p_1^{(1)},p_2^{(1)}$. 
Taking into account expression \eqref{prob_CBE} and using the estimated probabilities for each composite component in each group ($\hat{p}_1^{(0)},\hat{p}_2^{(0)},\hat{p}_1^{(1)},\hat{p}_2^{(1)}$) and the estimated pooled event probability of the composite endpoint ($\hat{p}_*$), the correlation is estimated by	
$$
\hat{\rho} = \frac{\hat{p}_* - \frac{\tilde{n}^{(0)}}{\tilde{n}}(1-\hat{q}_1^{(0)}\hat{q}_2^{(0)}) - \frac{\tilde{n}^{(1)}}{\tilde{n}}(1-\hat{q}_1^{(1)}\hat{q}_2^{(1)}) }{
	-\frac{\tilde{n}^{(0)}}{\tilde{n}}\sqrt{\hat{p}_1^{(0)}\hat{p}_2^{(0)}\hat{q}_1^{(0)}\hat{q}_2^{(0)}}
	-\frac{\tilde{n}^{(1)}}{n}\sqrt{\hat{p}_1^{(1)}\hat{p}_2^{(1)}\hat{q}_1^{(1)}\hat{q}_2^{(1)}}
}
$$
where $\hat{q}_k^{(i)}=1-\hat{p}_k^{(i)}$, and $\tilde{n}^{(i)}$ is the sample size in group $i$ in the blinded data. 
Based on these estimates we then compute the sample size ratio $d(\hat{p}_1^{(0)}, \hat{p}_2^{(0)},OR_1,OR_2,\hat{\rho})$ to select the endpoint.

\begin{figure}
	\centering
	\includegraphics[scale=0.7]{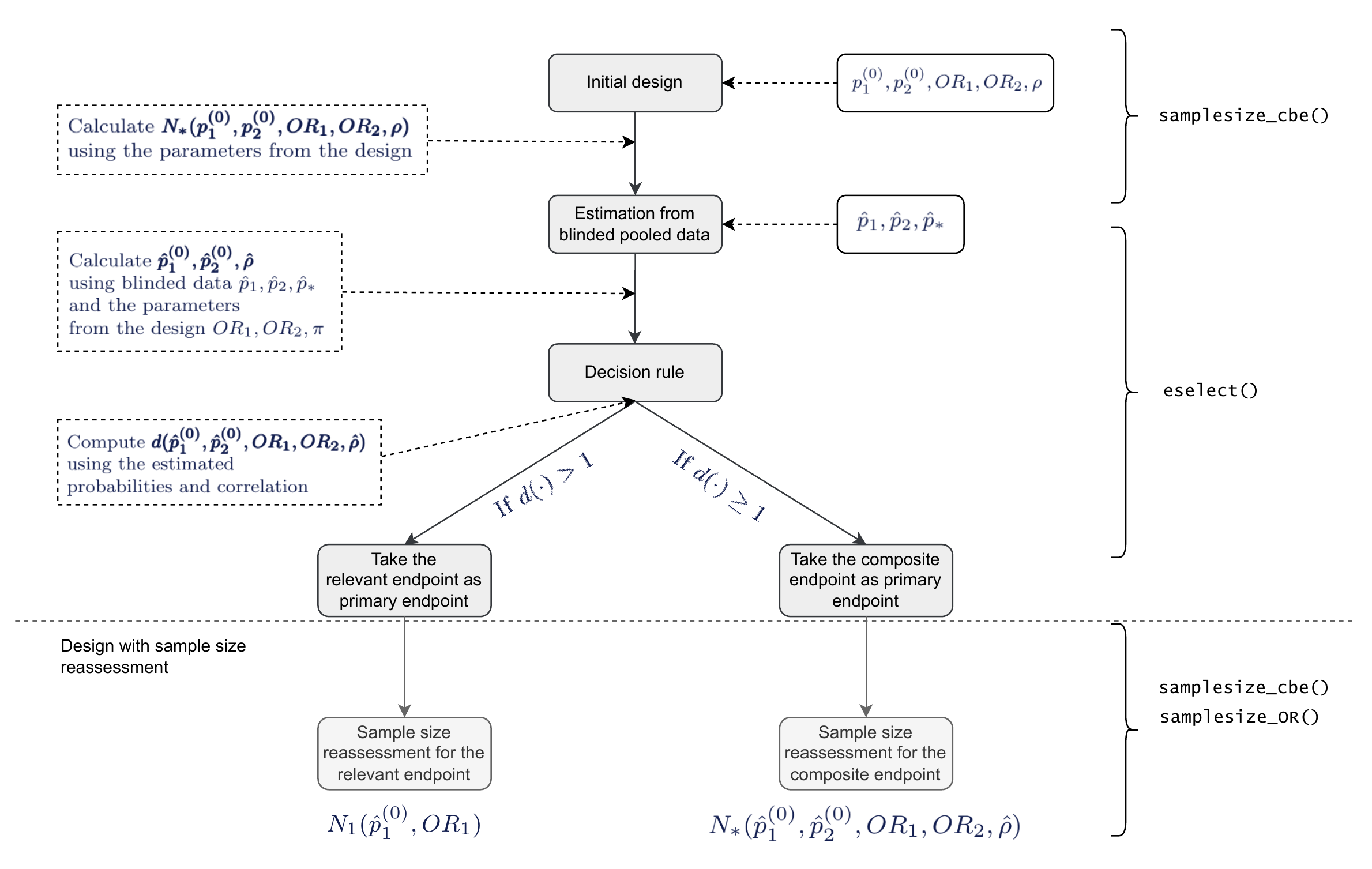} 
	\caption{Flow diagram of the adaptive design. The steps involved in adaptive design are illustrated in grey boxes. In the white boxes there are the necessary inputs, and explanations and outputs are in dotted white boxes. The R functions  to compute the corresponding steps are on the right side  
		(see Sect.  5 in the supplementary material). Here
		$p_1^{(0)},p_2^{(0)},OR_1,OR_2$ denote the design parameters for the endpoints $\varepsilon_1$ and $\varepsilon_2$ and $\rho$ is the correlation between $\varepsilon_1$ and $\varepsilon_2$ used for the calculation of the initial sample size, $n$;
		$\hat{p}_1^{(0)},\hat{p}_2^{(0)}$ denote the estimated event probabilities in control group for $\varepsilon_1$, $\varepsilon_2$ and $\varepsilon_*=\varepsilon_1\cup\varepsilon_2$ and $\hat{p}_k$ is the estimated pooled event probability of $\varepsilon_k$ ($k=1,2,*$) based on the blinded sample $n$; $N_1$ and $N_*$ denote the sample sizes for endpoint $\varepsilon_1$  and $\varepsilon_*$ (see Sect. \ref{design_CE} and \ref{design_RE}), respectively; and  $d(\cdot)$ is the decision function (see Sect. \ref{sect_adapdesign}).
	}%
	\label{scheme}%
\end{figure}

The diagram in Figure \ref{scheme} exemplifies the adaptive design if initially the composite endpoint is chosen as the primary endpoint. Note that in order to calculate the initial sample size for the composite endpoint, assumptions regarding the parameters' values determining the sample size have to be made. 

\subsection{Sample size reassessment} \label{sect_ssreassess}

After the endpoint has been selected based on the estimates  $\hat{p}_1^{(0)}, \hat{p}_2^{(0)},\hat{\rho}$,  evaluated from the blinded data, in addition the sample size can be recalculated.
When the composite endpoint is selected, the target sample size, computed from the above estimates and based on the pre-specified effect sizes $OR_1,OR_2$, is given by $N_*(\hat{p}_1^{(0)},\hat{p}_2^{(0)},OR_1,OR_2,\hat{\rho})$.  
Because the overall sample size cannot be smaller than the number of already recruited patients, the sample size  reassessment rule is given by  
$$n_a = \max\{\tilde{n}, N_*(\hat{p}_1^{(0)},\hat{p}_2^{(0)},OR_1,OR_2,\hat{\rho})\}$$
where $\tilde{n}$ denotes the number of patients recruited so far. 


If, in contrast,  the most relevant component is chosen as primary endpoint, the sample size can be reassessed to aim at a power of $1-\beta$ for this endpoint. The sample size calculation is based on the pre-specified effect size $OR_1$ and the estimated event probability $\hat{p}_1^{(0)}$. Thus, in this case the sample size reassessment rule is given by $n_a = \max\{\tilde{n}, N_1(\hat{p}_1^{(0)},OR_1)\}$.

If the selection is made at the interim analysis, $\tilde{n}<n$ and therefore the recalculation could result in a reduction of the  initially planned sample size. In contrast, if the selection is made at the planned end of the trial, $\tilde{n}=n$, the sample size can either remain unchanged or can be increased if required.

\subsection{Considerations for choosing the timing of the interim analysis}

As usual in adaptive trials, the timing of the interim analysis has to be fixed independently of the observed data and described in the trial protocol. For the proposed design, a reasonable strategy is to consider as initial sample size the minimum between the sample size for the relevant endpoint and the composite endpoint assuming a correlation of $0$, that is, 
$$
\tilde{n} = \min \{N_1(p_1^{(0)},OR_1), N_*(p_1^{(0)},p_2^{(0)},OR_1,OR_2,\rho=0) \}
$$ 
For correlation equals zero,  the required sample size for the composite endpoint is the smallest (assuming that only non-negatively correlated components are possible) \citep{Bofill2019}. Therefore, a reasonable strategy would be to fix the design as follows. First, conduct the selection of the endpoint based on blinded data after $\tilde{n}$ subjects. Then, reassess the sample size according to the rule defined in Section \ref{sect_ssreassess}. If the reassessed sample size is smaller than $\tilde{n}$, stop the trial and conduct the final (unblinded) analysis of the data. Otherwise, expand the trial with further subjects as needed and conduct the final (unblinded) analysis of the selected endpoint $n_a$. The maximum sample size is bounded by the maximum sample size coming from the sample size calculation for the relevant endpoint and composite endpoints assuming the largest possible correlation.


\section{Extension to more than two components and more than two arms}
\label{sect_extension} 

In this section, we address the recursive selection of the primary endpoint for more than two components, and discuss the extension to more than two arms. 

\subsection{Composite endpoints with more than two components}

Consider now a trial with $K$ potential endpoints of interest. We assume that they differ in importance and can be ordered according to their importance. 
Let $\varepsilon_1,\cdots,\varepsilon_K$ denote the endpoints ordered by decreasing importance. 
Let $p_k^{(0)}$ and $OR_k$ denote the event probabilities in the control group and the effect size for the endpoint $\varepsilon_k$ ($k=1, ...,K$). In the planning phase of the RCT, assumptions on the event probabilities, effect sizes, and correlations values are made to obtain an initial sample size estimate. 

The procedure to select the primary endpoint and recalculate the sample size accordingly for $K$ components is based on the following algorithm:
\begin{itemize}
	\item[] \textbf{Step 1:} Compare the required sample size for the endpoint $\varepsilon_1$ and the composite of the first and second endpoints, $\varepsilon_{*,2}= \varepsilon_1 \cup \varepsilon_2$ and compute the sample size ratio based on the estimated probabilities and assumed effect sizes, 
	$\hat{p}_1^{(0)},\hat{p}_2^{(0)},OR_1,OR_2$ and the estimated correlation between $\varepsilon_1$ and $\varepsilon_2$, denoted by $\hat{\rho}_{*,2}$.  If $d(\hat{p}_1^{(0)},OR_1,\hat{p}_2^{(0)},OR_2,\hat{\rho}_{*,2}) \geq 1$, then compute the event probability and effect size of the composite endpoint, $\varepsilon_{*,2}$, denoted by $\hat{p}_{*,2}^{(0)}$ and $OR_{*,2}$ and continue with the next step. Otherwise, select $\varepsilon_1$ and go to Step $K$. 
	
	\item[] \textbf{Steps $i=2,...,K-1$:} Compare the efficiency of using $\varepsilon_{*,i}$ over $\varepsilon_{*,i+1}= \varepsilon_{*,i} \cup \varepsilon_{i+1}$.
	
	Compute the sample size ratio based on
	$\hat{p}_{*,i}^{(0)}$, $OR_{*,i}$, computed in the previous step, and $\hat{p}_{i+1}^{(0)}$, $OR_{i+1}$, and the estimated correlation between $\varepsilon_{*,i}$ and $\varepsilon_{i+1}$, here denoted by $\hat{\rho}_{*,i+1}$.
	
	If $d(\hat{p}_{*,i}^{(0)},\hat{p}_{i+1}^{(0)},OR_{*,i},OR_{i+1},\rho_{*,i+1})\geq1$, then compute the parameters of the composite endpoint $\varepsilon_{*,i+1}$ and go to Step $i+1$. Otherwise, select $\varepsilon_{*,i}$ and go to Step $K$. 
	
	\item[] \textbf{Step $K$:} Reassess the sample size  based on the selected endpoint.
\end{itemize}

Using this recursive method, we only need the anticipated values of event probabilities in the control and effect sizes of the components ($\varepsilon_1,\cdots,\varepsilon_K$). If the composite endpoint is selected in the step $i$, this endpoint is considered as a component for the composite considered in the next step. For this reason, the corresponding parameters are recalculated and considered as anticipated values of the components in the next iteration. 

\subsection{Trials with more than two arms}

Consider a multi-armed RCT comparing the efficacy of $M$ treatments to a shared control treatment using the binary composite endpoint $\varepsilon_* = \varepsilon_1 \cup \varepsilon_2$. We test the $M$ individual null hypotheses $H_*^{(m)}: OR_*^{(m)} =1$ against the alternative $K_*^{(m)}: OR_*^{(m)} <1$ for each arm $m$ ($m=1, \cdots, M$), where $OR_*^{(m)}$ denotes the odds ratio for the composite endpoint in the $m$-th treatment arm.

Denoting the test statistics \eqref{test:ch} to compare treatment $m$ against control by $T_{*,n}^{(m)}$, as before we have that asymptotically $T_{*,n}^{(m)}\sim N(0,1)$. We reject the null hypothesis if $T_{*,n}^{(m)}<z_{\alpha/M}$, adjusting the threshold to account for the multiplicity of treatment arms. To size the trial, suppose that the expected effect size for the components is the same in all treatment arms, that is, $OR_k=OR_k^{(m)}$ for all $m$ ($k=1,2$). Additionally, as we did before, assume that the correlation between the components is equal across arms, $\rho=\rho^{(m)}$ for all $m$. Note that this implies that $OR_*=OR_*^{(m)}$ for all $m$. 
For each individual comparison, the sample size is 
$N_*(p_*^{(0)},OR_*)$ as described in Section 2, and as the trial considers a shared control the total sample size for the trial is:
$$
N_{*,M}(p_*^{(0)},OR_*) = N_*(p_*^{(0)},OR_*) \cdot (M- (M-1)\cdot \pi)
$$ 
where $\pi$ is the allocation proportion to the control group. 
The sample size for the multi-armed RCT can then be determined by means of the same set of parameters $(p_1^{(0)},p_2^{(0)},OR_1,OR_2,\rho)$.

For the most relevant endpoint, the null and alternative hypotheses for treatment $m$ are $H_1^{(m)}: OR_1^{(m)} =1$ and $K_1^{(m)}: OR_1^{(m)} <1$. Consider the test statistics $T_{1,n}^{(m)}$ to compare treatment $m$ against control, which is asymptotically $N(0,1)$ under $H_1^{(m)}$ and reject $H_1^{(m)}$ if  $T_{1,n}^{(m)}<z_{\alpha/M}$. Assuming the   effect sizes to be equal across arms, $OR_1=OR_1^{(m)}$, the total sample size for the trial would be 
$N_{1,M}(p_1^{(0)},OR_1) = N_1(p_1^{(0)},OR_1) \cdot (M- (M-1)\cdot \pi)$, 
where $N_1(p_1^{(0)},OR_1)$ is the required sample size for each individual comparison. 

The sample size ratio $d(\cdot)$ is then reduced to the same as used in \eqref{decision}, and  the adaptive design proposed in Section \ref{sect_adapdesign} can then be applied analogously as for the case of a two-armed trial.  Hence if  $d(\cdot)>1$, the design for testing the efficacy using the most relevant endpoint(s) is chosen, otherwise the composite endpoint is, and in either case we recalculate the sample size  using the event probability and the correlation estimates. 
As the same effects are assumed for all arms, the same procedure can also be used to estimate the probabilities under the treatment group and the correlation. This assumption allows the estimates to be blinded and permits the selection of the primary endpoint to be the same for all arms. However, if we relax these assumptions, it could result in different selection strategies, e.g., maximizing the minimum power across all arms or partly unblinding the data (blind pooling treatment data if arms are not finishing at the same time like in multi-arm platform trials).

\section{Motivating example 
	in Peritoneal Dialysis Trials}
\label{sect_example}


Consider a trial in peritoneal dialysis with the primary endpoint major adverse peritoneal events (MAPE), defined as the composite endpoint of peritonitis and peritoneal membrane deterioration ($\varepsilon_1$) and 
technical failure ($\varepsilon_2$).  
It has to be noted the MAPE initially consists of three components, but we grouped peritonitis and peritoneal events together for the sake of illustration.  Also, suppose that the endpoint of peritonitis and peritoneal membrane deterioration can be considered as the most relevant endpoint that could serve as sole primary endpoint. Table \ref{Table:endpoints} summarizes the considered endpoints. 

\cite{Boehm2019} reported event probabilities of the individual endpoints and combinations thereof. 
We use these estimated event probabilities as estimates for the event probabilities in the control group at the design stage of the trial (see Table \ref{Table:endpoints}). We discuss  
the efficiency of using MAPE ($\varepsilon_*=\varepsilon_1 \cup \varepsilon_2$) over the endpoint of peritonitis and peritoneal membrane deterioration ($\varepsilon_1$) alone, and illustrate the  design with adaptive selection of the primary endpoint at the interim analysis and sample size reassessment.

\begin{figure}
	\centering
	\subfloat[Initial sample size when using the trial design with the most relevant endpoint of peritonitis and peritoneal membrane deterioration (RD), or with the composite endpoint Major Adverse Peritoneal Events (CD).]{\includegraphics[scale=0.7]{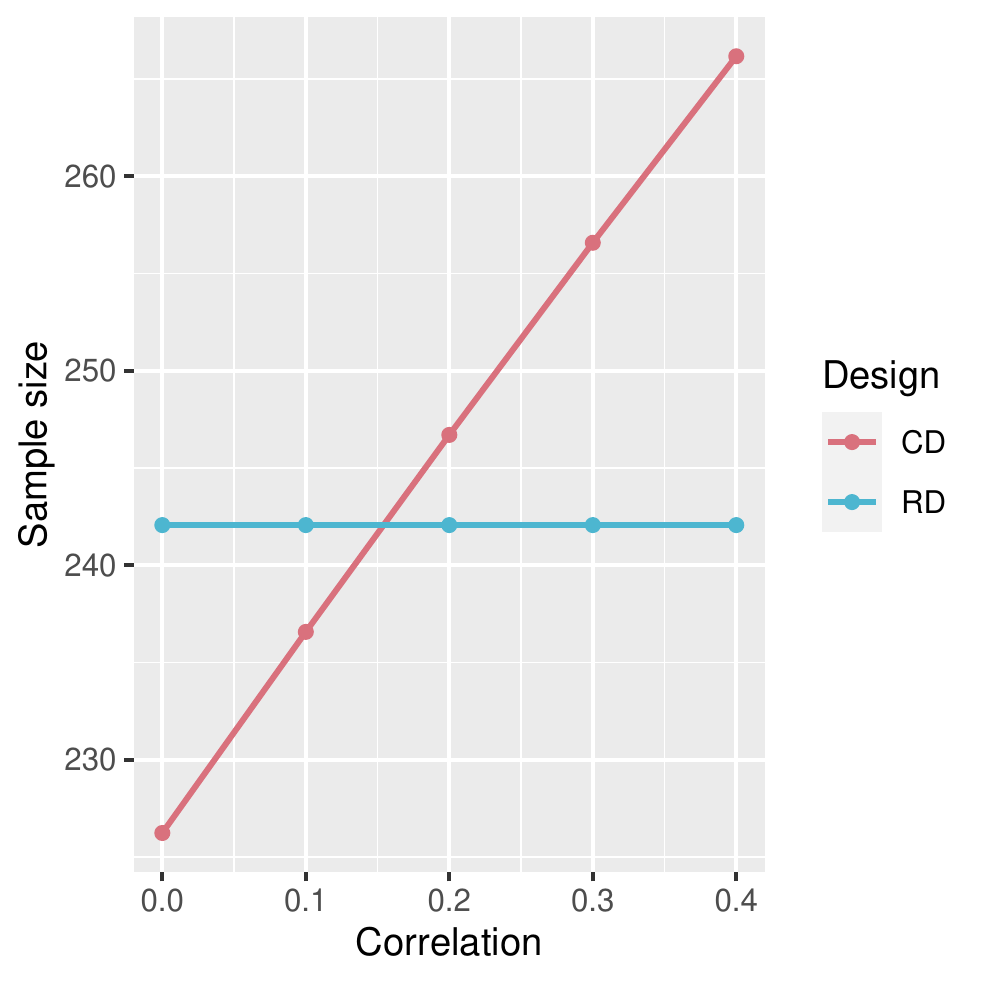}}%
	\qquad
	\subfloat[Power when using the fixed design with the most  relevant endpoint of peritonitis and peritoneal membrane deterioration (RD), fixed design with the composite endpoint Major Adverse Peritoneal Events (CD), and the adaptive design (AD)]{\includegraphics[scale=0.7]{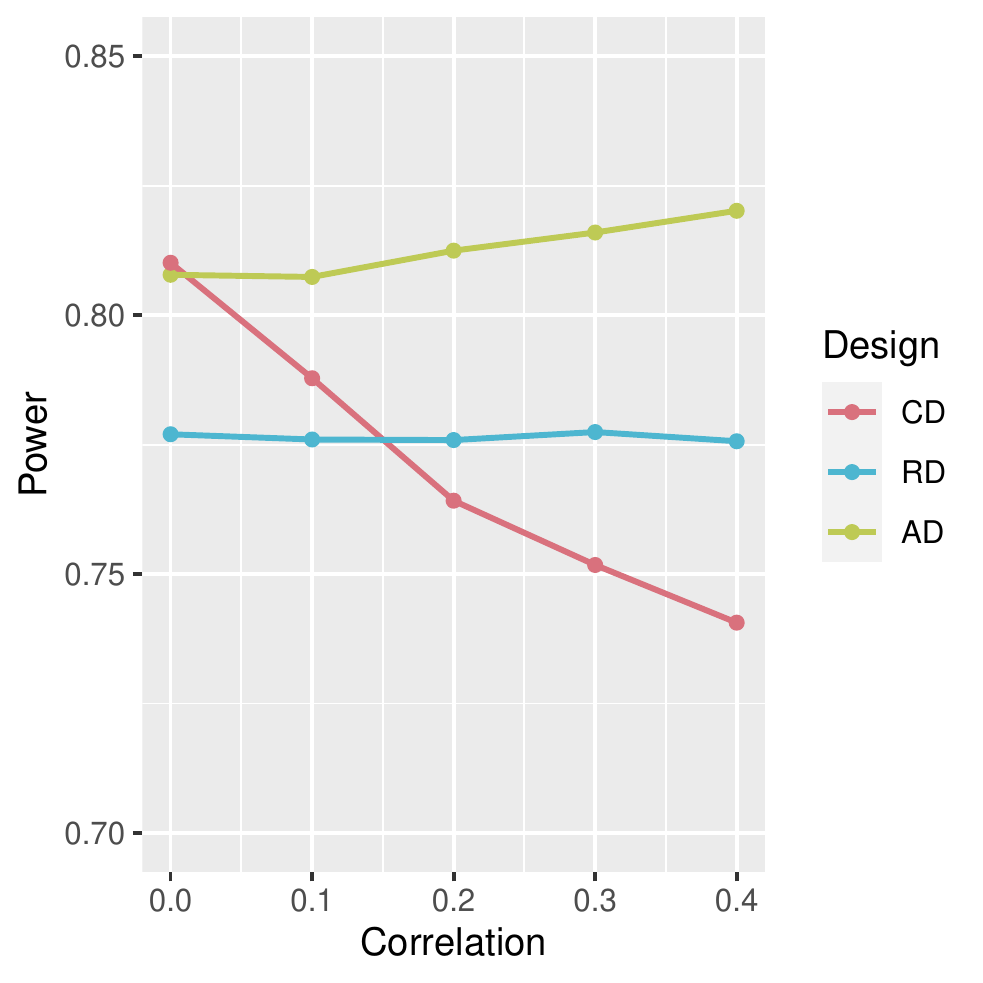}}%
	\caption{Sample size and power depending on the design and the correlation between the endpoint of peritonitis and peritoneal membrane deterioration ($\varepsilon_1$) and technical failure ($\varepsilon_2$).}%
	\label{example_ss_power}%
\end{figure}

In Figure \ref{example_ss_power} (a), we depict the sample size required for MAPE with respect to the correlation between $\varepsilon_1$ and $\varepsilon_2$, and the sample size if only using $\varepsilon_1$, both based on the parameters assumed at the design stage (Table \ref{Table:endpoints}). 
We can observe that the sample size increases with respect to the correlation.   
In Figure \ref{example_ss_power} (b), we show the power of the trial when using a fixed design with the endpoint  MAPE, $\varepsilon_*$, as primary endpoint, assuming that the correlation equals 0, a fixed design with the most relevant endpoint $\varepsilon_1$, and when using the proposed adaptive design. We notice that the adaptive design allows to maintain the power of the trial at 0.80 and is superior to the power obtained when using the fixed design.
The decision rule of the adaptive design is such that it selects the endpoint that requires the smallest estimated sample size. Furthermore, if this sample size does not result in the desired power, it is readjusted based on information from the interim analysis. So when the estimated correlation is lower than 0.2, the adaptive design typically selects the composite endpoint as primary endpoint and recomputes the sample size using the estimated correlation. When the estimated correlation is larger or equal than 0.2, then  the most relevant endpoint is selected and the sample size is reassessed accordingly.

\section{Simulation study}
\label{sect_sim} 

\subsection{Design and main assumptions}

We simulate the statistical power and significance level under different scenarios and consider two-arm RCTs with two binary endpoints and parameters as given in Table \ref{Table_scenarios}.
The correlation between the endpoints is assumed to be equal for both groups. Since the range of possible correlations depends on ($p_1^{(0)},p_2^{(0)},OR_1,OR_2$), scenarios in which the correlation is not within the valid range are discarded. 

We compare the actual type 1 error rate and power of the proposed adaptive design with fixed designs using the relevant or composite endpoints as primary endpoint. Specifically, we consider the following designs: 
\begin{itemize}
	\item Adaptive design: trial design whose primary endpoint is adaptively selected between the composite and the most relevant endpoint based on blinded data; 
	
	\item Composite endpoint design: trial design without adaptive modification of the primary endpoint.
	The primary endpoint is the composite of $\varepsilon_1$ and $\varepsilon_2$.
	
	\item Relevant endpoint design: trial design without adaptive modification of the primary endpoint. The primary endpoint is the most relevant endpoint ($\varepsilon_1$);
	
\end{itemize}
We differentiate between two types of designs: those with selection of the components of the composite endpoint at the end of the study and those with selection at interim analysis.
In the first, the selection is based on blinded data at the pre-planned end of the trial, using the total sample size planned at the design stage. 
In the second, we select the primary endpoint based on blinded information obtained at an interim analysis after $50\%$ of the observations are available. 
For these designs, we consider designs with and without sample size recalculation after the interim analysis.

In trials with endpoint selection at the end of the study or at interim but without recalculation of sample size, the planned sample size $n$ is calculated to have $0.80$ power to detect an effect of $OR_1$ on the most relevant endpoint at significance level $\alpha = 0.05$. We use this sample size for the three designs being compared. Therefore, the composite endpoint in this case is intended to be used only if it leads to an increase in power to the study. 
On the other hand,  the (initial) sample size $n$ for those trials with sample size reassessment is calculated to have $0.80$ power to detect an effect of $OR_*$ on the composite endpoint at significance level $\alpha = 0.05$, where $p_*^{(0)},OR_*$ used for the sample size calculations are computed based on the components' parameters ($p_1^{(0)},p_2^{(0)},OR_1,OR_2$) and assuming correlation equal 0. 
Therefore, in this case, the adaptive design serves to readjust the values anticipated in the design for the composite endpoint if the components are correlated, and to compare the efficiency of the design compared to its most relevant component, and thus to change the primary endpoint if the composite endpoint is less efficient. We summarize in Table \ref{Table_designs} the trial designs considered for the simulation study.

For each combination ($p_1^{(0)},p_2^{(0)},OR_1,OR_2,\rho$), we simulated $100000$ trials of size $n$ according to each design (adaptive design, composite endpoint design, and relevant endpoint design).
To evaluate the power, we considered the alternative hypothesis $H_1$ in which $OR_1,OR_2<1$ (and therefore $OR_*<1$). We simulated based on the values assumed in the design for $OR_1$, $OR_2$ and the resulting $OR_*$ computed based on the parameters ($p_1^{(0)},p_2^{(0)},OR_1,OR_2,\rho$).
To evaluate the type 1 error rate, 
the same set of scenarios were considered as for the power in terms of the values used for the sample size calculation
but we simulated under 
the global null hypothesis $H_0$ so $OR_1=OR_2=1$ (and therefore $OR_*=1$).  The total number of scenarios is $1166$.


\subsection{Selection at the end of the trial}

\begin{figure}%
	\centering
	\subfloat[Without sample size reassessment]{\includegraphics[scale=0.7]{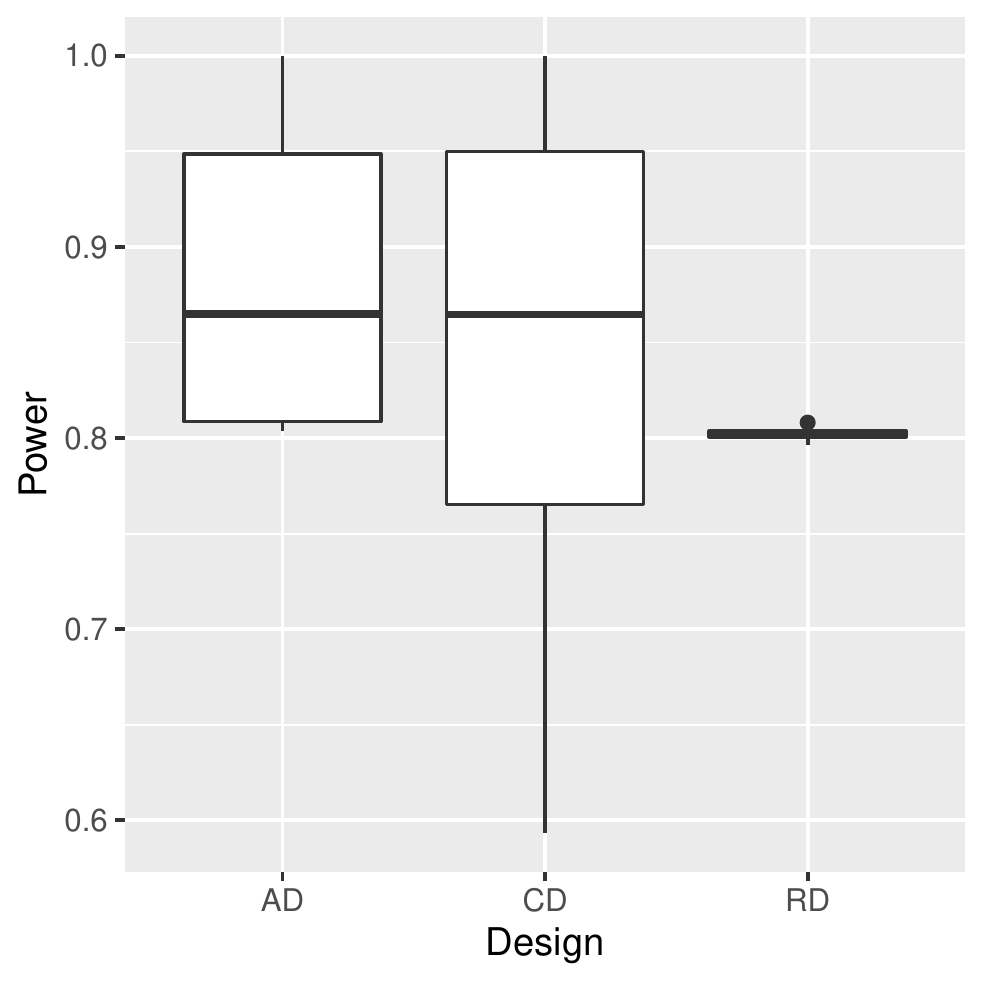}}%
	\qquad
	\subfloat[With sample size reassessment]{\includegraphics[scale=0.7]{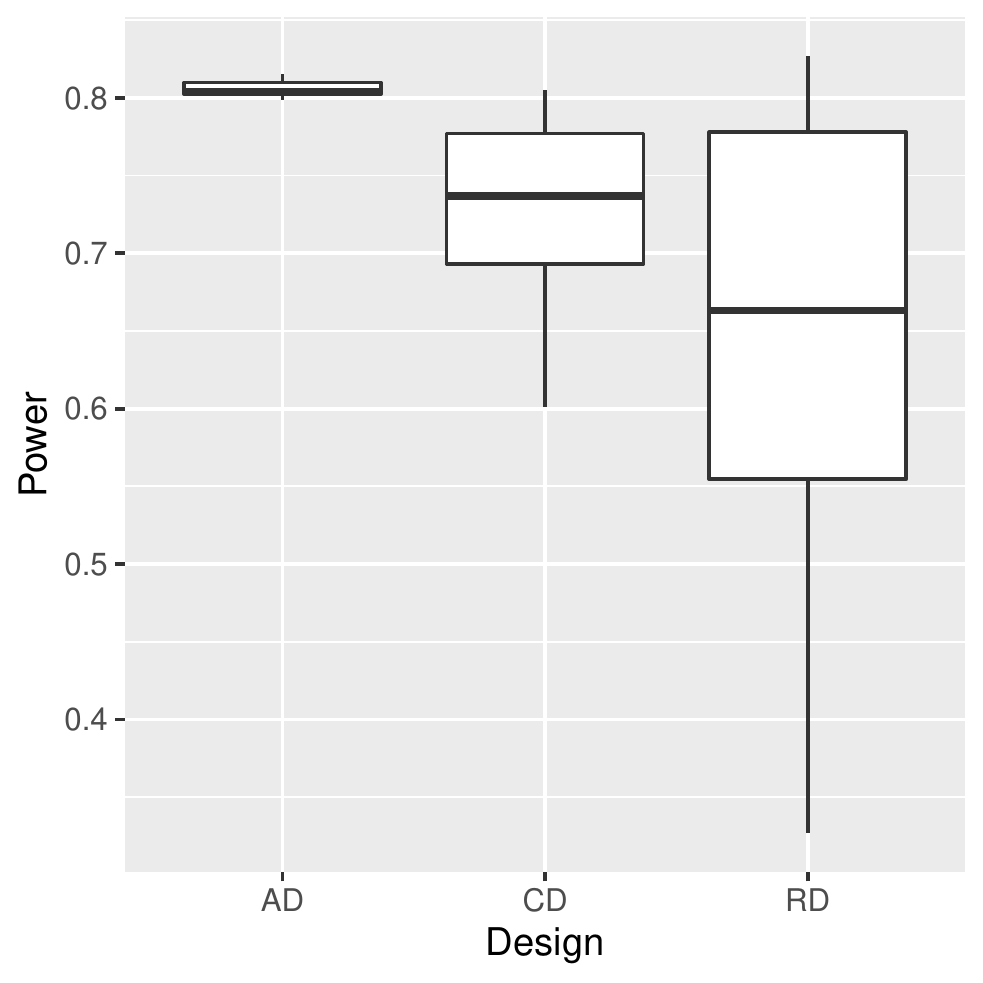}}%
	\caption{Power under composite design (CD), relevant design (RD) and adaptive design (AD). Case (a) refers to designs with selection at the end of the trial without sample size recalculation for which the initial sample size was computed to have $0.80$ power to detect effects on the relevant endpoint.  Case (b) refers to those designs with selection at the interim analysis and with sample size recalculation for which the sample size was computed the have $0.80$ power to detect effects on the composite endpoint assuming zero correlation between the components.} 
	\label{results_boxplot}%
\end{figure}

As expected, for the scenarios under the alternative hypotheses the powers when using the most relevant endpoint have mean $0.80$, as the sample sizes were calculated for this endpoint. The powers when using the composite endpoint range from $0.60$ to $1.00$ with mean $0.85$. With the adaptive design, the powers take values between $0.80$ and $1.00$, and have mean $0.88$. Results are summarized in Figure \ref{results_boxplot}.

\begin{figure}%
	\centering
	\subfloat[Without sample size reassessment]{\includegraphics[width=1\linewidth]{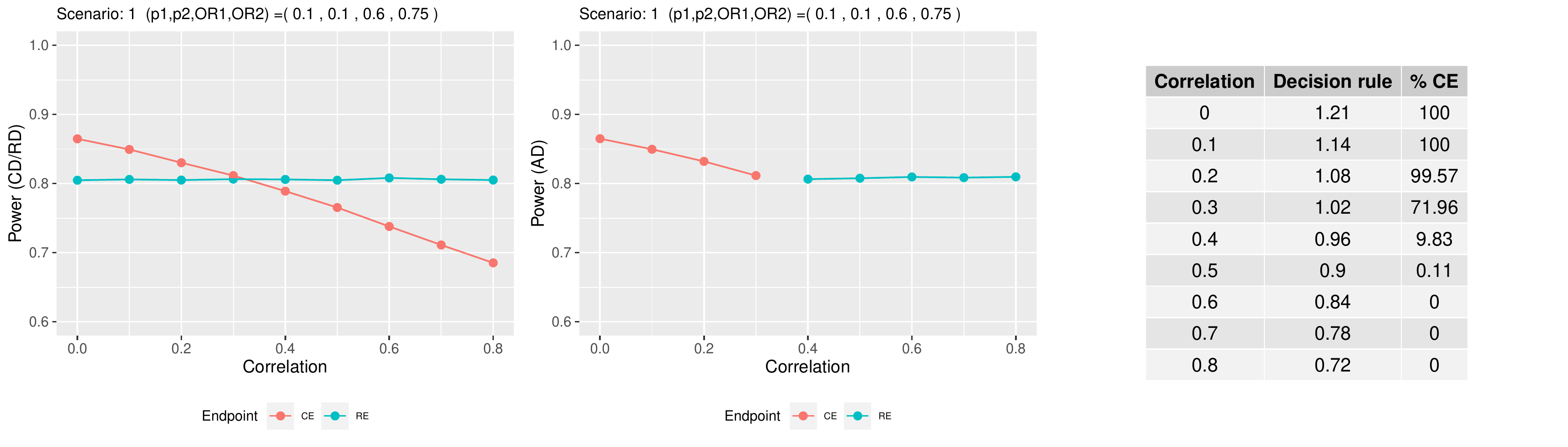}}%
	\qquad
	\subfloat[With sample size reassessment]{\includegraphics[width=1\linewidth]{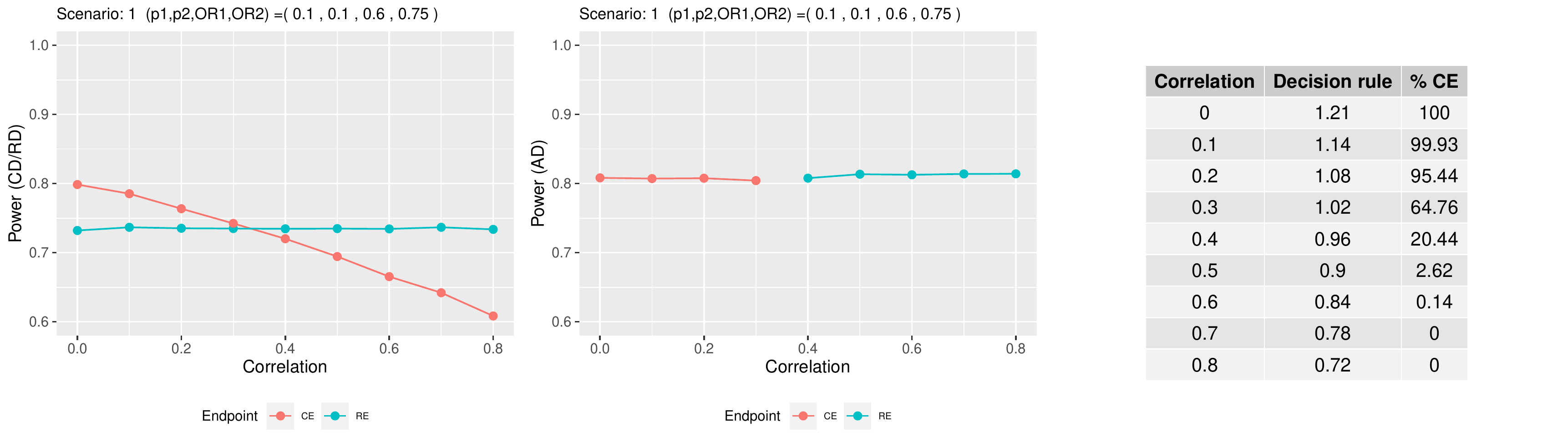}} 
	\caption{Power under composite design (CD), relevant design (RD) and adaptive design (AD) with respect to the correlation between the components. 
		In (a), trials are initially sized to detect an effect on the relevant endpoint and the ADs select the primary endpoint at the end of the trial; in (b), trials are sized to detect an effect on the composite endpoint and ADs select the primary endpoint at the end of the trial and subsequently recalculate the sample size.
		Tables on the right side shows the value of the decision rule computed using the parameters' values used for the simulation and the percentage of cases in which the composite endpoint is selected as the primary endpoint.
		Note that for the CD and RD, the primary endpoint is the composite endpoint (CE) and relevant endpoint (RE), respectively, for the AD, the primary endpoint changes depending on the correlation. } 
	\label{fig:examplepowers_rss}%
\end{figure}

To illustrate the properties of the adaptive design, consider a specific scenario (see Figure \ref{fig:examplepowers_rss}). For a given combination of combination of ($p_1^{(0)},p_2^{(0)},OR_1,OR_2$), we plot the empirical power for each design (adaptive design, composite endpoint design, and relevant endpoint design) for different correlations $\rho$. 
The colors in the power plots indicate which endpoint is optimal for the given parameters $p_1^{(0)},p_2^{(0)},OR_1,OR_2,\rho$.
From there, 
we observe that when the power for the composite endpoint design is  greater than $0.80$ regardless of the correlation value, the decision in the adaptive design is to use the composite. Likewise, if the composite design's power is less than $0.80$, the relevant design will be chosen.
Also note that the decision rule, i.e., the ratio of sample sizes in \eqref{decision}, decreases with respect to the correlation. This is due to the sample size for composite endpoints increasing as the components are more correlated. Indeed, for a given set of marginal parameters ($p_1^{(0)},p_2^{(0)},OR_1,OR_2$), the composite design is more efficient the lower the correlation. Therefore, when using the adaptive design, the decision rule chooses the composite endpoint when the estimated correlation between the components is small and chooses the most relevant endpoint when the estimated power using the composite falls below $0.80$. Thus, the power of the adaptive design is always greater than $0.80$. 
In the supplementary material we plot  the empirical power for each design as function of the correlation $\rho$ for all scenarios considered in the simulation. 
For the scenarios simulated under the global null hypothesis (i.e., $OR_*=OR_1=OR_2=1$), all designs control the type 1 error rate at the nominal level $\alpha=0.05$.

\subsection{Selection at the interim analysis}

\subsubsection{With sample size reassessment}
The initial sample size in these settings was computed to detect an effect on the composite endpoint, assuming uncorrelated components ($\rho=0$).
For the relevant endpoint design, the powers in this case range from $0.33$ to $0.85$ with mean $0.64$; and when using the composite endpoint range from $0.60$ to $0.80$ with mean $0.72$. For the adaptive design, in contrast, the powers have mean $0.80$ (see Figure \ref{results_boxplot}). The proposed adaptive design, therefore, ensures that the target power is achieved, either by keeping the composite endpoint as primary but correcting the correlation value assumed in the design and recalculating the sample size accordingly in the interim analysis, or by modifying the endpoint to the most relevant endpoint and adjusting the corresponding sample size. 
To illustrate the properties of the adaptive design we again focus on a selected scenario (see Figure \ref{fig:examplepowers_rss}). For the other considered cases see the supplementary material. We observe that when using the adaptive design, the power is always maintained at 0.80, while for the composite endpoint design it depends on the true value of the correlation and the extent to which it deviates from the correlation assumed at the design stage (which is, in our case, $\rho=0$).
On the other hand, the type I error rate is as well maintained at $0.05$.

\subsubsection{Without sample size reassessment} 

When using the adaptive design with endpoint selection at an interim analysis without sample size reassessment, the observed results are slightly worse  to those obtained when selecting the endpoint at the end of the study as the estimates have a higher variability. 
The type 1 error rate under the null scenarios investigated is again well controlled (data not shown).

\subsection{Additional considerations}

\subsubsection{Comparison between blinded and unblinded estimators}
In this work, we proposed an adaptive modification of the primary endpoint and sample size reassessment based on parameter estimates, estimated from the blinded (interim) data. Alternatively, the event probabilities in the control group and the correlation between endpoints can be estimated using the unblinded data (but still using the a priori estimates of the effect sizes). To assess the properties of this alternative approach, we simulated adaptive trials for the above scenarios with selection at the interim analysis or at the end of the trial, and without sample size assessment. 
The power of the adaptive design using unblinded data is equal to or slightly higher than when using blinded data (see the supplementary material). However, when evaluating the type 1 error we observe that when unblinded information is used there is an inflation of type 1 error when using a conventional frequentist test as defined in Section \ref{sect_intro}. If the selection should be done on unblinded data in an interim analysis more complex adaptive closed testing strategies \citep{bauer2016twenty} have to be used and the data cannot naively be pooled over stages.

\subsubsection{Properties of the design if there is no treatment effect in some of the components}
We additionally assessed the power of the designs  in scenarios where i) there is no effect in the most relevant endpoint; and ii) there is no effect in the additional endpoint.  
In these settings the adaptive design is not the most powerful design: the power of the adaptive design is between the power using only the relevant and the composite endpoints (see the supplementary material).

\section{Discussion}
\label{sec7}
In this paper, we proposed an adaptive design that allows the modification of the primary endpoint based on blinded interim data and recalculates the sample size accordingly. The design selects either a composite endpoint or the endpoint with the most relevant component as the primary endpoint, based on the ratio of sample sizes needed in the corresponding designs  to achieve a certain power. This ratio depends on the event probabilities in the control group and the effect sizes for each composite component, and the correlation between them.  
We presented estimators for the event probabilities and correlation based on blinded data obtained at an interim or the pre-planned final analysis and proposed to use them to compute the sample size ratio. The advantage of using blinded data is that the type 1 error rate is not inflated when performing the conventional frequentist tests for the selected primary endpoint at the end of the trial. In all null scenarios investigated no substantial inflation of the type 1 error could be observed. This was expected as both the selection and sample size reassessment were based on blinded data \citep{posch2018estimation, kieser2003simple} and not the observed treatment effect directly. The results obtained from the proposed adaptive design are, therefore, in line with the requirements of regulatory agencies for adaptive designs with endpoint selection \citep{FDA_2019}, since the adaptation rules for blinded endpoint selection are predefined in the design and the methods considered keep the type 1 error control. 

If the selection is done at the end, we showed that the proposed design is more powerful than the fixed designs using the composite endpoint or its more relevant component as the primary endpoint in all scenarios considered in the simulation study. The simulations have shown that as long as the marginal effect sizes have been correctly specified, the power never falls below the nominal power. In addition, a re-estimation of the sample size has been proposed by adjusting the sample size at the interim stage to incorporate the estimated correlation and estimated event probabilities in the control group based on the assumed effect sizes.  Since the correlation between the components is rarely known and therefore not usually taken into account when sizing a trial with composite endpoints, we want to emphasize that this sample size calculation could be useful even without adaptive modification of the primary endpoint. As in trials with composite endpoints, the required sample size increases as the correlation increases, we proposed to start the trial assuming correlation equals zero and recalculate the sample size accordingly based on the blinded data.
If sample size reassessment is not considered, then the best results are achieved when the selection of the primary endpoint is made at the end of the study due to the smaller variability of the blind estimates. However, for consistency checks and to convince external parties such as regulators, it might be reassuring to have a second independent sample, that has not been used before to determine the endpoint.

We focused on the estimation of the correlation based on blinded data but also considered estimators based on unblinded data (see the supplementary material). We compared the operating characteristics of trial designs using blinded and unblinded correlation estimators. Power is slightly higher when using the unblinded estimator. However, it may lead to a substantial type 1 error inflation. Throughout this work, in both blinded and unblinded data cases, we assumed that correlations are equal across treatment groups. This assumption, although common, may in some cases not be satisfied. We discuss the implications of this assumption in terms of the design and interpretation, also an approach to tailor the proposed design to cases where the correlations are not equal in the supplementary material. 
To allow for unequal correlations and blinded selection, one has to fix the effect size not only for the components, but also for the composite endpoint. There is a trade-off by having fewer assumptions but more fixed design parameters.
However, further empirical investigations are needed to evaluate how plausible it is that the equal correlation across arms assumption will not be met and the impact of different correlations on interpreting the effect of the composite endpoint. 

In this paper, we consider trials with large sample sizes, so derivations of sample size calculations are based on asymptotic results. In the case of trials with small sample sizes, it should be noted that smaller sizes would result in lower precision in event estimates, which could affect the variable decision and sample size recalculation. 
Finally, we extended the proposed design for trials with more than two groups and more than two components. Further extensions can be considered by giving greater flexibility in terms of the selection of the primary endpoint (e.g. choosing different primary endpoints according to treatment arm) and considering platform designs where the treatment arms enter and leave at different times during the trial (and therefore interim analysis also at different times). Extensions to complex designs such as those mentioned above and designs with time-to-event endpoints are open to future research.

\section{Supplementary Material}

Supplementary material includes further derivations, discussion on  extensions for unequal correlations across arms, introduction of other association measures,  an overview of the R package,  an additional example based on a conducted randomized trial in cardiology including the R code, and other results from the simulation study.
The R code to reproduce the results of this article is available at \url{https://github.com/MartaBofillRoig/eselect}.

\section*{Acknowledgments}

We thank the reviewers and associate editor for the comments and suggestions that helped  improve the manuscript.
M. Bofill Roig and G. G\'{o}mez Melis were partly supported by the Ministerio de Ciencia e Innovaci\'{o}n (Spain) under Grant PID2019-104830RB-I00; the Departament d'Empresa i Coneixement de la Generalitat de Catalunya (Spain)  under 2017 SGR 622 (GRBIO). 
M. Bofill Roig, F. Koenig and M. Posch are members of the EU Patient-centric clinical trial platform (EU-PEARL). EU-PEARL has received funding from the Innovative Medicines Initiative 2 Joint Undertaking under grant agreement No 853966. This Joint Undertaking receives support from the European Union's Horizon 2020 research and innovation programme and EFPIA and Children's Tumor Foundation, Global Alliance for TB Drug Development non-profit organization, Spring- works Therapeutics Inc. This publication reflects the author's views. Neither IMI nor the European Union, EFPIA, or any Associated Partners are responsible for any use that may be made of the information contained herein.

{\it Conflict of Interest}: None declared.


\bibliography{refs}

\newpage


\begin{table}[!h]
	\caption{Endpoints in peritoneal dialysis. Event probability and odds ratios for peritonities and peritoneal membrane deterioration and Technical failure endpoints. Event probability and odds ratio for MAPE endpoint computed assuming zero-correlation between the components of the composite endpoint. \label{Table:endpoints}}
	{\tabcolsep=4.25pt
		\begin{tabular}{@{}ccccc@{}}
			\hline
			&Endpoint & Event probability & Odds ratio \\\hline
			\textbf{Individual} & Peritonities and peritoneal membrane deterioration ($\varepsilon_1$) & $0.615$ & $0.52$ \\ 
			\textbf{endpoints:} & Technical failure ($\varepsilon_2$) & $0.15$ & $0.66$ \\ \hline
			\textbf{Composite}  &Major Adverse Peritoneal Events
			(MAPE) & $0.703$ &  $0.50$
			\\ 
			\textbf{endpoint:} &($\varepsilon_1 \cup \varepsilon_2$) &   &  \\
			\hline
	\end{tabular}}
\end{table}

\begin{table}[!h]
	\caption{Settings for the simulation: ($p_1^{(0)},p_2^{(0)},OR_1,OR_2$) denote the parameters for the endpoints $\varepsilon_1$ and $\varepsilon_2$, $\rho$ is the correlation between the endpoints,  $\omega$ is the percentage of initial sample size used for the estimation and decision rule computation, and $\alpha$ and $1-\beta$ refer to the significance level and power. \label{Table_scenarios}}
	{\tabcolsep=4.25pt
		\begin{tabular}{@{}ccccc@{}}
			\hline
			Endpoints & & &  \\ \hline
			$\varepsilon_1$ & $p_1^{(0)}$ & $0.1, 0.2$ &
			$OR_1$ & $0.6, 0.8, 1$ \\
			$\varepsilon_2$ & $p_2^{(0)}$ & $0.1, 0.25$ &
			$OR_2$ & $0.75, 0.8, 1$ \\[3pt]
			& $\rho$ & $0, 0.1, 0.2, 0.3, 0.4, 0.5, 0.6, 0.7, 0.8$ & & \\ \hline
			\textbf{Design} & &  &   \\\hline
			& $\omega$ & $0.5$, $1.0$; \\ 
			& $\alpha$ & $0.05$ & \\
			& $1-\beta$ & $0.80$  \\
			\hline
	\end{tabular}}
\end{table}

\begin{table}[!h]
	\caption{Outline of trial designs considered for the simulation, including: the sample size specification for the initial calculation, whether it was based on relevant endpoint (RE) or composite endpoint (CE); and, in the case of the adaptive design, at which point in the trial the endpoint selection is made and whether sample size recalculation is considered.  \label{Table_designs}}
	{\tabcolsep=4.25pt
		\begin{tabular}{@{}ccc@{}}
			\hline
			All designs & Adaptive design &  \\[3pt] \cline{2-3}
			\textbf{Initial} & \textbf{Endpoint selection} & \textbf{Sample size} \\
			\textbf{Sample size} & & \textbf{reassessment} \\ [3pt] \hline
			Based on RE ($OR_1$) & At the end of the study & No \\[3pt] \hline 
			Based on CE ($OR_*$ with $\rho=0$)  & \multirow{2}{*}{At the interim analysis} & Yes\\ [3pt]
			Based on RE ($OR_1$) &  & No \\  
			\hline
	\end{tabular}}
\end{table}

\newpage

\includepdf[pages=-]{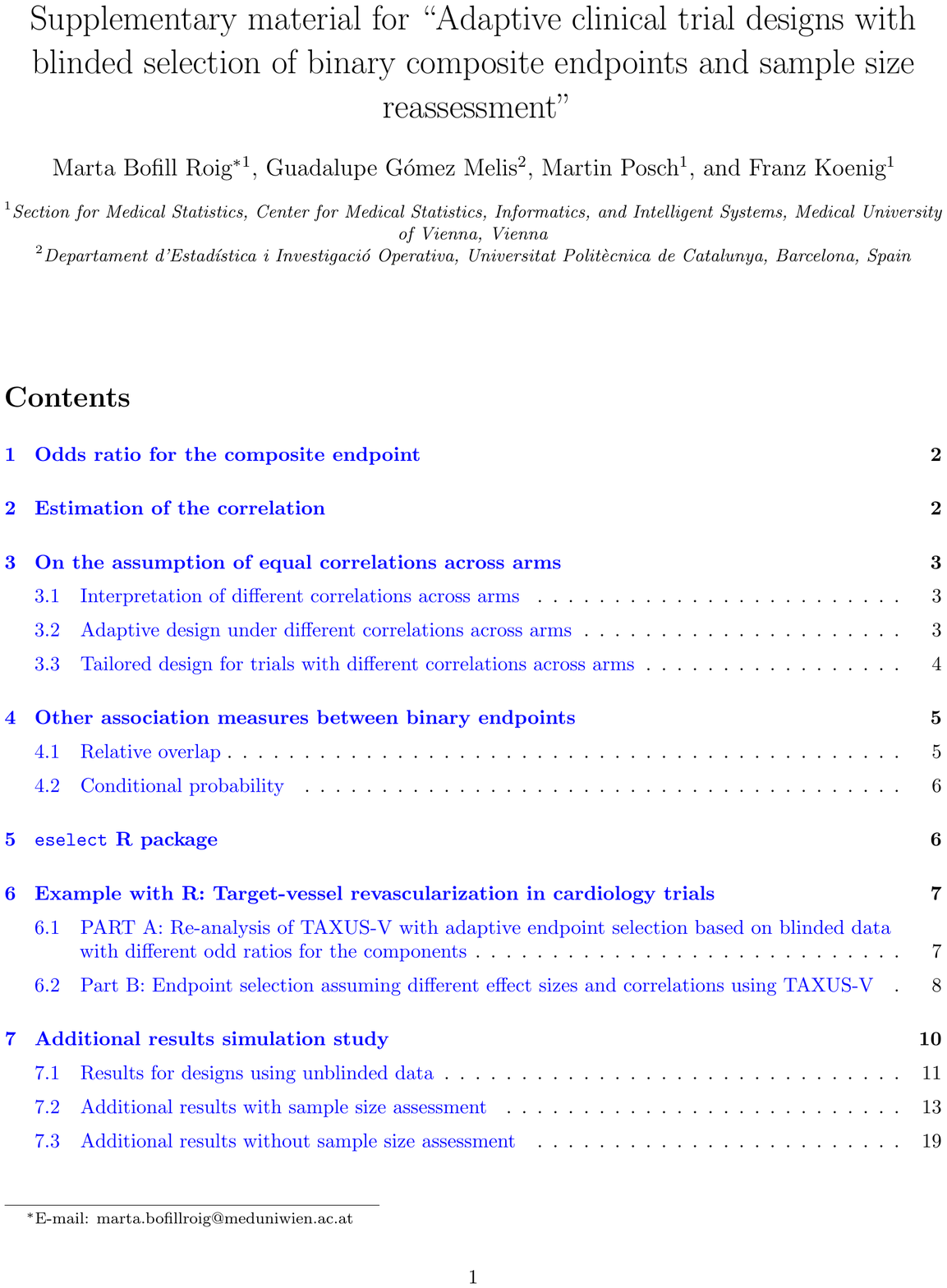}

\clearpage

\end{document}